\documentclass[10pt,letterpaper]{article}

\usepackage{cogsci}
\usepackage{booktabs}
\cogscifinalcopy 

\usepackage{pslatex}
\usepackage{apacite}
\usepackage{float} 
\usepackage{graphicx}
\usepackage{subcaption}
\usepackage{enumitem}

\title{Model Human Learners:\\Computational Models to Guide Instructional Design}
 
\author{{\large \bf Christopher J. MacLellan (cmaclell@gatech.edu)} \\
  School of Interactive Computing\\
  Georgia Institute of Technology\\Atlanta, GA 30308 USA}

\begin{document}

\maketitle

\begin{abstract}
Instructional designers face an overwhelming array of design choices, making it challenging to identify the most effective interventions.
To address this issue, I propose the concept of a {\it Model Human Learner}, a unified computational model of learning that can aid designers in evaluating candidate interventions.
This paper presents the first successful demonstration of this concept, showing that a computational model can accurately predict the outcomes of two human A/B experiments---one testing a problem sequencing intervention and the other testing an item design intervention.
It also demonstrates that such a model can generate learning curves without requiring human data and provide theoretical insights into why an instructional intervention is effective.
These findings lay the groundwork for future Model Human Learners that integrate cognitive and learning theories to support instructional design across diverse tasks and interventions.


\textbf{Keywords:} 
Artificial Intelligence; Education; Learning; Skill acquisition and learning; Symbolic computational modeling
\end{abstract}

\section{Introduction}
In their seminal paper, \citeA{Card:1986vx} propose the concept of a {\it Model Human Processor}---a unified information-processing model that codifies psychological theories to predict human performance in user interfaces.
They argue that interface designers could use such a model to make more informed design decisions that improve usability.
I propose an analogous concept: a {\it Model Human Learner}---a unified computational model that codifies cognitive and learning theories to predict human learning in instructional systems. 
I argue that instructional designers could use this kind of model to make more informed instructional design decisions that improve pedagogical effectiveness.

An analysis by \citeA{Koedinger:2013hp} suggests that instructional designers face a crisis of choice, as design spaces are combinatorial and contain trillions of configurations.
How can instructional designer determine the best choices?
Currently, they rely on A/B experiments to navigate this vast space, but human studies are costly and time consuming.
Also, each experiment provides only a single bit of information, so instructional designers are essentially playing a losing game of 20 questions with nature \cite{Newell:1973tt}. 

Computational models of learning, such as those proposed by \citeA{Maclellan:2017thesis} and \citeA{weitekamp2023computational}, offer a potential solution.
Unlike abstract, mathematical models of learning, such as the Additive Factors Model \cite{Cen:2009ve} or Bayesian Knowledge Tracing \cite{Corbett:1994ux}, which fit functions to performance data, computational models of learning are {\it mechanistic}. 
Similar to cognitive architectures \cite{Langley:2008ka}, these models simulate how knowledge is updated in response to practice and how performance evolves over time.
I argue that these models can realize the Model Human Learner concept, letting instructional designers simulate A/B experiments to test alternative interventions and identify the most promising ones before conducting costly human studies.

Prior research has explored several applications of computational models of learning.
\citeA{Li:2013vd} investigated their use for discovering cognitive models, while
\citeA{Matsuda:2011wf} explored how they can promote ``learning-by-teaching,'' where students learn by instructing simulated students.
\citeA{maclellan2022domain} and \citeA{weitekamp2020interaction,weitekamp2024ai2t} examined their applications for tutor authoring.
Others have studied their utility for theory testing \cite{rachatasumrit2023content,Lee:2009ty,NanLi:2010wi}.

Researchers have also proposed using these models to predict the effects of different instructional design choices \cite{MacLellan:2016tqa,maclellan2023optimizing}.
While these studies generate reasonable predictions, {\bf they have not yet been validated against human data.}
Some prior work has examined model-human alignment \cite{rachatasumrit2023content,Maclellan:2017thesis,weitekamp2019toward,weitekamp2020investigating}, but this has focused primarily on theory testing, rather than instructional design evaluation.
This paper addresses that gap by comparing model predictions to outcomes from two human A/B experiments.
Specifically, I demonstrate that computational models of learning can:
\begin{itemize}[itemsep=-3pt, topsep=3pt]
    \item Accurately predict the main effects of two human experiments---one evaluating a problem sequencing intervention and the other testing an item design intervention;
    \item Generate learning curve predictions that closely match human trends, without training on human data first; and
    \item Offer theoretical insights into why specific interventions work, challenging a prior  hypothesis by \citeA{Lee:2015gb} and suggesting a novel explanation.
\end{itemize}

\section{The Computational Model}
This study employs a computational model from the {\it Apprentice Learner Architecture} \cite{Maclellan:2017thesis,weitekamp2020interaction,maclellan2022domain}, which integrates mechanisms from prior models of human learning, including ACM \cite{LangIey:1984uh}, CASCADE \cite{VanLehn:1991we}, STEPS \cite{Ur:1995wr}, and SimStudent \cite{Li:2013vd}. 

Specifically, I use the {\it Trestle} model,\footnote{For a full description, which is beyond the scope of the current paper, see \citeA{maclellan2022domain}.} which provides an account for how skills are incrementally acquired from a mixed combination of worked examples and correctness feedback.
When presented with a problem, such as the fraction arithmetic problem shown in Figure~\ref{fig:fractions-tutor}, the model matches previously learned skills against the current state. If any skills match, it executes the one with the highest utility to generate a step. If no skills match, which is common early in learning, then the model requests a demonstration from the tutor. For example, in Figure~\ref{fig:fractions-tutor}, the tutor might demonstrate placing a 6 in the lower left denominator conversion box. The model, equipped with basic arithmetic primitives for adding, subtracting, multiplying, and dividing, searches for a sequence of mental operations (i.e., a procedure) that explains this demonstration. 
It might identify that multiplying the two given fraction denominators  produces the demonstrated value.
The agent then generalizes this procedure, removing specific values to form a reusable skill.
It then applies separate learning mechanisms to identify the conditions for applying the skill and to update its utility.
When the agent uses this skill on subsequent problems and receives feedback from the tutor, it further refines the skill's conditions and utility.



\section{Study 1: Fraction Arithmetic Tutor}

\subsection{Human Data}

\begin{figure}[t] 
    \includegraphics[width=0.47\textwidth]{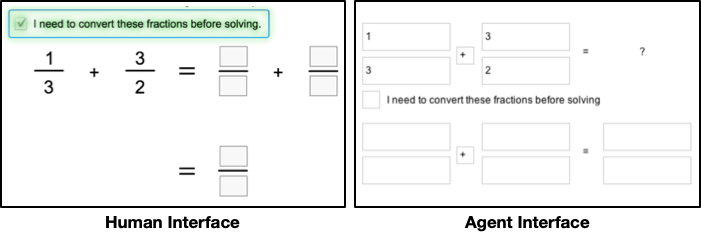}
    \caption{Fraction arithmetic tutor interfaces.} 
    \label{fig:fractions-tutor}
\end{figure}

To evaluate this model's ability to predict human learning outcomes, I used data from the {\it Fraction Addition and Multiplication} dataset accessed via DataShop \cite{Koedinger:2010tj}.\footnote{https://pslcdatashop.web.cmu.edu/DatasetInfo?datasetId=1190} \citeA{Patel:ho4PITKp} collected these data during an experiment to examine whether blocking or interleaving fraction arithmetic problems leads to better learning.

The dataset includes tutor data from 79 students solving 24 fraction addition problems (ten with same denominators and 14 with different denominators) and 24 fraction multiplication problems using the tutor interface shown in Figure~\ref{fig:fractions-tutor}. 
Students had to check the ``I need to convert these fractions before solving'' box to reveal the conversion fields. For fraction addition requiring conversion, only the cross multiplication strategy (multiplying denominators) was accepted; alternative strategies were marked incorrect.
After training, students took a posttest within the tutor, which consisted of four fraction multiplication and four fraction addition problems (two with same denominators and two with different). No hints or feedback were provided during the posttest.

Students were randomly assigned to one of two conditions. In the blocked condition, problems were presented in three sequential blocks: fraction addition (same denominators), fraction addition (different denominators), and finally, fraction multiplication. The order of the problems within each block was randomized. In the interleaved condition, problems were presented in a fully randomized order. 
This experiment aimed to test whether interleaving enhances learning compared to the blocking approach recommended by the Common Core State Standards. \citeA{Patel:ho4PITKp} found that students in the blocked condition performed better during training, while those in the interleaved condition performed better on the posttest. This suggests that although interleaved practice results in errors during training, it ultimately leads to greater long-term learning. The goal of this study is to determine whether the computational model can accurately predict this main experimental effect.

\begin{figure}[t] 
    \centering
    \begin{subfigure}[b]{0.47\textwidth} 
        \includegraphics[width=\textwidth]{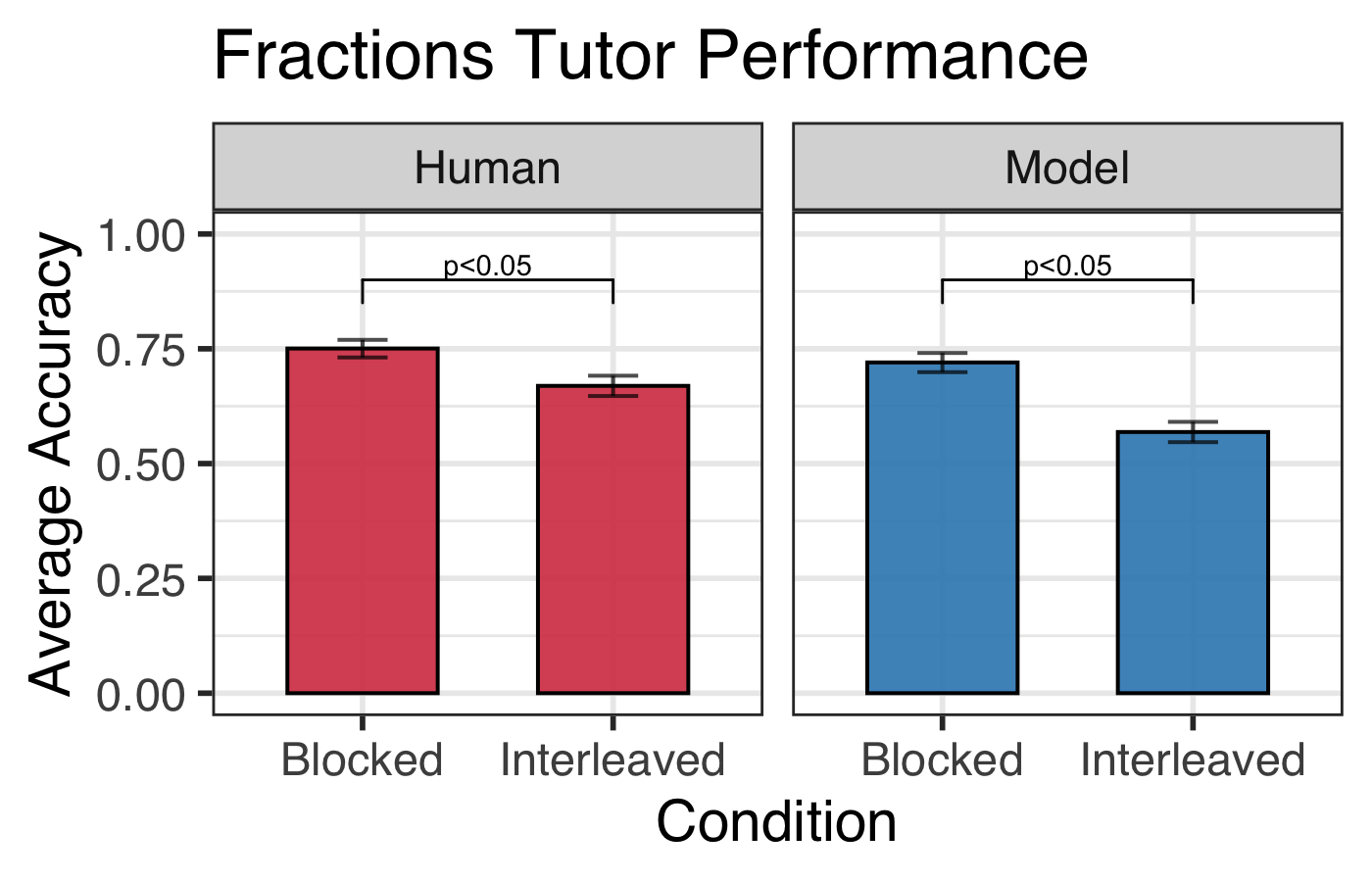}
    \end{subfigure}
    \hfill 
    \begin{subfigure}[b]{0.47\textwidth} 
        \includegraphics[width=\textwidth]{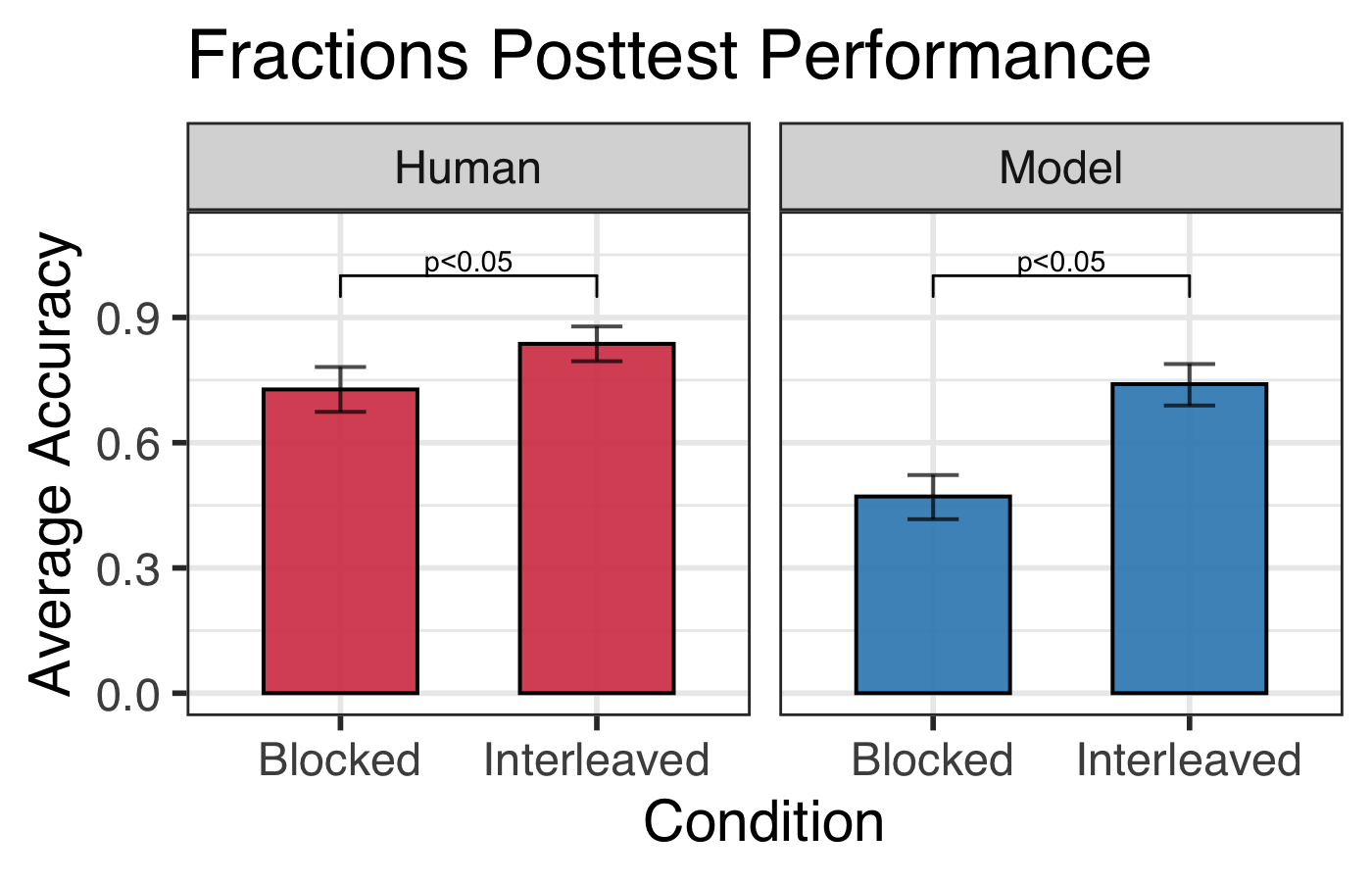}
    \end{subfigure}
    \caption{Fraction arithmetic accuracy on tutor and posttest problems with 95\% confidence intervals (CIs).} 
    \label{fig:fractions-overall}
\end{figure}

\subsection{Simulation and Analysis Methods}
To simulate human behavior, I created an instance of the model (an agent) for each student and connected it to the machine-readable version of the tutor, shown on the right in Figure \ref{fig:fractions-tutor}.
This isomorphic tutor, developed using Cognitive Tutor Authoring Tools \cite{aleven2006cognitive}, provided each agent with the same sequence of problems that its corresponding human student received. Importantly, agents were not constrained to take the same actions as their human counterparts. Thus, a tutor was necessary to provide the agents with appropriate hints and feedback during training, as the log data alone was insufficient for conducting the simulation. 

\begin{figure}[t] 
    \centering
    \begin{subfigure}[b]{0.47\textwidth} 
        \includegraphics[width=\textwidth]{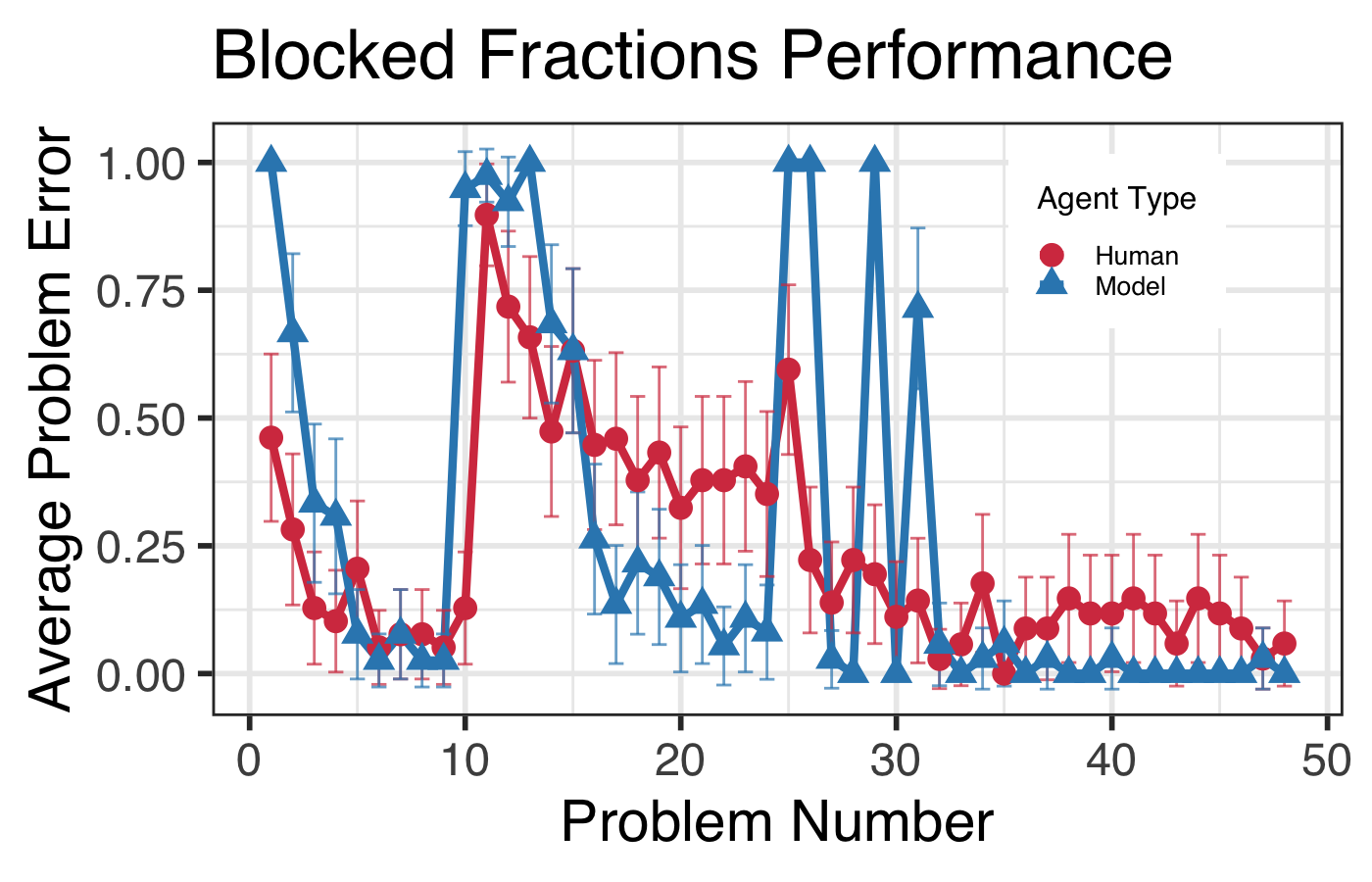}
    \end{subfigure}
    \hfill 
    \begin{subfigure}[b]{0.47\textwidth} 
        \includegraphics[width=\textwidth]{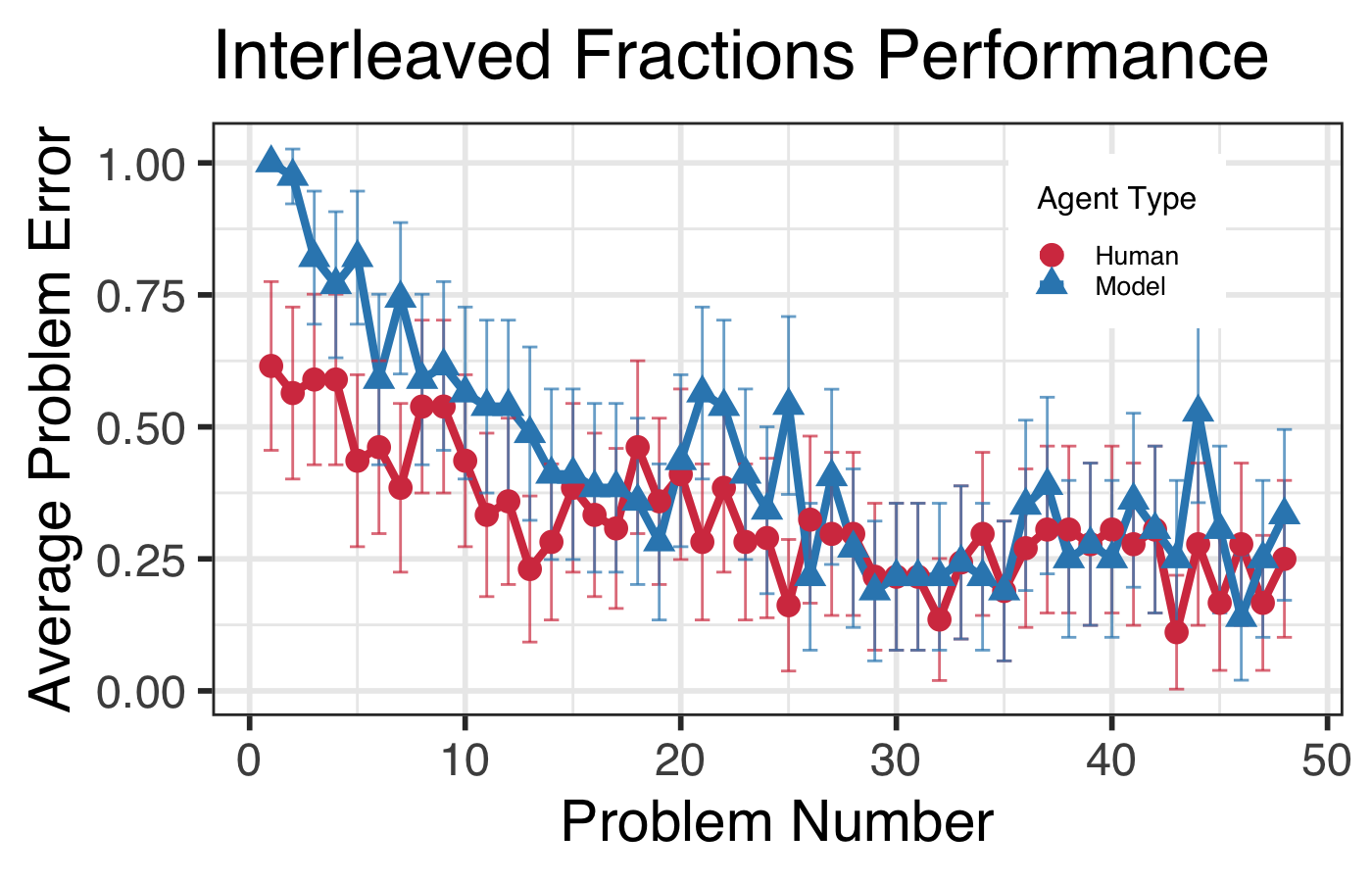}
    \end{subfigure}
    \caption{Fraction arithmetic learning curves with 95\% CIs. The blocked tutor transitions types at problems 11 and 25.} 
    \label{fig:fractions-learning-curve}
\end{figure}

One key technical limitation of the isomorphic tutor was that it did not support hidden fields. In the human tutor, the fields for converting fractions remained hidden until the student checked the ``I need to convert these fractions before solving'' box, whereas in the machine-readable tutor, these fields were always visible. As a result, simulated students could make errors---such as attempting to convert fractions before selecting the conversion box---that were impossible in the human tutor. However, preliminary simulations showed that agents rarely made such errors.

After training, the human students took a posttest to assess their fraction arithmetic knowledge. This assessment was administered directly within the tutor with hints and correctness feedback disabled. Students could enter any values in the text fields and they could check the conversion box at any time, which revealed the conversion fields regardless of whether conversion was necessary. At any point, they could press the done button and advance to the next problem. \citeA{Patel:ho4PITKp} evaluated students based on the percentage of problems solved correctly. The simulated agents also completed the posttest within the tutor. During the posttest, hints and feedback were disabled. Agents were marked as correct only if they completed all steps of a problem correctly, in which case they advanced to the next problem. If they made any mistakes or requested a hint, the problem was immediately marked incorrect and they were moved to the next problem. 

To statistically evaluate the effect of instructional condition on both humans and agents, I conducted mixed-effect logistic regression analyses. 
During data preparation, I identified that seven students had corrupted posttest data and one additional student had corrupted tutor and posttest data.
After removing these cases, the dataset included tutor data from 78 students and posttest data for 71 students.
The agent simulations were based on the tutor data, so they produced tutor and posttest predictions for the 78 students with complete tutor data.

To assess tutor performance, I modeled correctness as the dependent variable, including fixed effects for condition, problem type (add same, add different, and multiply), and problem count (number of prior problems of the same type). I also included an interaction between problem type and count to capture different learning rates for problem types. For human data, I incorporated a random intercept for each student to account for individual differences. Since all agents had identical prior fraction knowledge (none), no random effect was included. For posttest performance, I again modeled problem correctness as the dependent variable, with instructional condition and problem type as fixed effects and a random effect for student. Problem count was excluded, as students did not receive any feedback during testing, so learning was not expected. Unlike in training, agents had varying prior knowledge at posttest based on their training, so the regression included a random effect for student.

\begin{table*}[ht]
\small
\centering
\caption{Mixed-effect regression models fit to humans and agents fraction arithmetic tutor and posttest data.}
\begin{tabular}{l|c|c||c|c}
\toprule
\textbf{Fixed Effects} & \textbf{Human Tutor} & \textbf{Model Tutor} & \textbf{Human Posttest} & \textbf{Model Posttest}\\ 
\midrule
Intercept (Add Diff) & 0.35 [0.21, 0.57]* & 0.12 [0.08, 0.17]* & 1.43 [0.73, 2.81] & 0.12 [0.06, 0.22]* \\
Condition (Interleaved) & 0.50 [0.28, 0.89]* & 0.40 [0.34, 0.47]* & 2.35 [1.0, 5.52]* & 7.75 [4.06, 14.78]* \\
Type: Add Same & 4.54 [2.76, 7.45]* & 5.59 [3.48, 9.04]* & 4.42 [2.28, 8.59]* & 1.43 [0.82, 2.48] \\
Type: Mult & 10.39 [6.92, 15.62]* & 7.49 [4.99, 11.37]* & 4.23 [2.42, 7.40]* & 38.08 [19.73, 73.47]* \\
Count & 1.16 [1.12, 1.20]* & 1.39 [1.33, 1.45]* & --- & --- \\
Type: Add Same × Count & 1.17 [1.07, 1.27]* & 0.93 [0.87, 1.00] & --- & --- \\
Type: Mult × Count & 0.94 [0.91, 0.98]* & 0.85 [0.81, 0.89]* & --- & --- \\
\toprule
\textbf{Random Effects} &  & & & \\ 
\midrule
$\sigma^2$ & 3.29 & --- & 3.29 & 3.29 \\ 
$\tau_{00}$ (Student) & 1.47 & --- & 2.00 & 0.79 \\ 
Intraclass Correlation & 0.31 & --- & 0.38 & 0.19 \\
N (Student) & 78 & --- & 71 & 78 \\
\toprule
Observations & 3559 & 3559 & 567 & 624 \\ 
Marginal/Conditional $R^2$ & 0.254 / 0.484 & --- & 0.100 / 0.441 & 0.499/0.596\\ 
$R^2$ (Tjur) & --- & 0.278 & --- & --- \\ 
\bottomrule
\end{tabular}\\
\textit{Note.} The fixed effect entries are have the format: Odds Ratio [95\% Confidence Interval].
* denotes $p < .05$.
\label{tab:fractions_regression}
\end{table*}

\subsection{Results}
\subsubsection{Main Effect of Condition}
The overall tutor and posttest performance for both humans and agents are shown in Figure~\ref{fig:fractions-overall}.
The results of the regression analysis are shown in Table~\ref{tab:fractions_regression}.
Mirroring the prior findings of \citeA{Patel:ho4PITKp}, I find that while humans have lower tutor performance in the interleaved condition (odds ratio: 0.50, $p<0.05$), 
they have higher posttest performance (odds ratio: 2.35, $p<0.05$). 
My analyses shows that agents exhibit this same effect.
Agents in the interleaved condition have lower tutor performance (odds ratio: 0.40, $p<0.05$), but higher posttest performance (odds ratio: 7.75, $p<0.05$).
Beyond the main effect of condition, the model exhibits most of the other effects, and direction of effects, that are present in the human data.
There are two exceptions: the effect of practice on the add same problem performance and the intercept on the posttest. 
However, in both situations either the human or model regression is not significant, suggesting that more data is needed. 



\subsubsection{Learning Curves}

I plotted the errors on each problem, averaging across students within condition. 
The resulting learning curves are shown in Figure~\ref{fig:fractions-learning-curve}.
Unlike models that generate predictions by fitting a function to the student data, such as the Additive Factors Model \cite{Cen:2006th,MacLellan:2015to}, these learning curves are {\it parameter-free} \cite{weitekamp2019toward} predictions based solely on task structure.
There is a large difference between the models and humans on earlier opportunities.
This suggests many students have prior fraction knowledge---their average first problem error was 53.8\%.
In contrast, the agents have zero prior fraction knowledge---they have an average first problem error of 100\%.
Despite this difference, the model does surprisingly well at {\it qualitatively} capturing the main effects.
For example, both humans and agents exhibit high error rates when transitioning from one problem type to another in the blocked condition and have similar asymptotic error.

\subsection{Discussion}
These results show that the model can successfully predict the main experimental effects of a fraction arithmetic problem ordering manipulation.
They also show that the model can generate reasonable parameter-free learning curve predictions. These results are a clear example of how tutor A/B experiment results can be predicted in a completely theory-driven way using a computational model of learning.

One caveat is that the model only qualitatively predicts the experimental effects. It does not accurately predict the absolute tutor and posttest scores for each student (or their average) because it does not currently account for prior fractions knowledge.
This is particularly noticeable in earlier practice opportunities, where the model predicts that human performance should be much worse than what is observed.
As previously mentioned, these differences are due to prior fraction arithmetic skills that students bring to the tutoring system, which are not accounted for in the current model. The model is essentially predicting what human performance would look like if the students did not have any prior fraction arithmetic skills. While these exaggerated error rates might be useful for detecting transitions between skills, such as when using learning curve analysis to develop knowledge-component models \cite{Corbett:1994ux}, they also suggest an opportunity for improvement. 

Some researchers have started to explore different ways to account for this knowledge.
\citeA{weitekamp2019toward} propose statistically estimating how much
previous practice each student has had with each type of problem and pretraining agents on a comparable number of problems to initialize prior knowledge.
Alternatively, \citeA{maclellan2023optimizing} suggest iteratively adjusting each agent's prior fraction knowledge to minimize the differences between agent and human learning trajectories.
Both approaches require human data, limiting their use when testing novel designs;
more research is needed to better account for prior knowledge in this case.

\section{Study 2: Box and Arrows Tutor}

\subsection{Human Data}

As a second test of whether a computational model can  predict the results of a human experiment, I used the {\it Box and Arrow Tutor Data (Turk Study)} dataset accessed via DataShop.\footnote{https://pslcdatashop.web.cmu.edu/DatasetInfo?datasetId=1123}
\citeA{Lee:2015gb} collected these data from Mechanical Turk as part of an experiment to investigate how different instructional choices effect student learning of problem-solving rules.
This experiment used the box and arrows tutor shown in Figure~\ref{fig:box-and-arrows-tutor}, which presents learners with two types of problems: (1) easy problems, where students enter the solution to the arithmetic problem from the first row into the empty box in the second row (e.g., $22/11=2$ in the right example from the figure), and (2) hard problems, where the student enters a value in second row that makes the lower arithmetic problem equal the value the base of the arrow points to (e.g., $7-4 = 3$ in the left example from the figure).
The easy problems were designed to be obvious, leveraging students' prior knowledge, whereas the hard problems were designed to be non-obvious, requiring students to discover the rule.

The data was collected from 202 students solving 32 box and arrow problems (16 easy and 16 hard).
The main experimental manipulation explored whether students perform better when receiving constrained or unconstrained problems. The constrained problems were designed so only the correct procedure produced a whole-number answer; other procedures (e.g., applying the easy procedure to a hard problems) produced a fractional answer. In contrast, unconstrained problems produced whole-number answers for incorrect procedures too. Students received no explicit instruction, but they did receive correctness feedback on answers.

\citeA{Lee:2015gb} frame learning on this task as a search process where students generate and test rules to see if they produce correct results.
They also hypothesized that constrained problems bias students towards the correct procedure because they make the correct procedure easier to compute than the incorrect procedure (working with whole numbers is easier than fractional numbers).
They found that students in the constrained condition had lower error rates on both easy and hard problems than those in the unconstrained condition.

My model is a direct instantiation of \citeA{Lee:2015gb}'s theory of learning; mainly, agents learn by generating and testing rules, making use of those that work and discontinuing use of those that do not.
However, working with whole and fractional numbers are equally difficult for the agents, so we can use them to directly test \citeA{Lee:2015gb}'s hypotheses for why constrained problems aid learning.
If the agents exhibit the same main effect as the humans, then it provides further evidence that computational models of learning can predict experimental outcomes.
Such a finding would, however, be direct evidence against \citeA{Lee:2015gb}'s hypothesis for why constrained problems make learning easier.
 
\subsection{Simulation and Analysis Method}

\begin{figure}[t] 
    \includegraphics[width=0.47\textwidth]{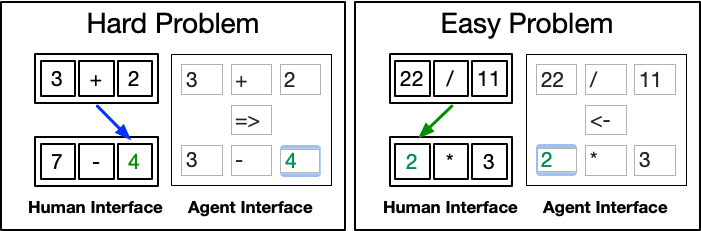}
    \caption{Box and arrows tutor interfaces.} 
    \label{fig:box-and-arrows-tutor}
\end{figure}

I used the same simulation procedure as study 1, creating an agent for each student and connecting it to a machine-readable version of the box and arrows tutor, shown in Figure~\ref{fig:box-and-arrows-tutor} to simulate learning.
An analysis of the human data showed that the majority of human students solved the easy problems correctly on the first attempt. In contrast, only a few students got the  hard problems correct on the first try because they are specifically designed to prevent prior knowledge use. Thus, for my analysis, I trained the agents on all the problems, but only analyzed their performance on the hard problems, because my model does not take into account prior knowledge, which seems to play a major role in solving easy problems.\footnote{Half of the simulated agents received prior training on 16 easy problems; I pre-trained the agents in case there was positive or negative transfer from the easy problems to the hard ones.} The focus on hard problems makes it possible to compare agents and humans in a situation where they have matched prior knowledge. This subset of the human data contained 202 students solving 16 hard problems, for a total of 3,232 observations.

To evaluate the effect of instructional condition, I conducted a mixed-effect logistic regression analysis similar to the one used for the fractions task. I modeled problem correctness as the dependent variable, including fixed effects for condition and problem count and a random effect for student. Unlike the fractions analysis, this regression model did not include an effect for problem type because all problems were of of the same type (hard). This mixed-effect regression model has a close correspondence to the ANOVA analysis conducted by \citeA{Lee:2015gb}. When applying this regression analysis to the agents, the random effect for student was removed because the agents all have identical initial conditions.

\subsection{Result}

\begin{figure}[t] 
    \includegraphics[width=0.47\textwidth]{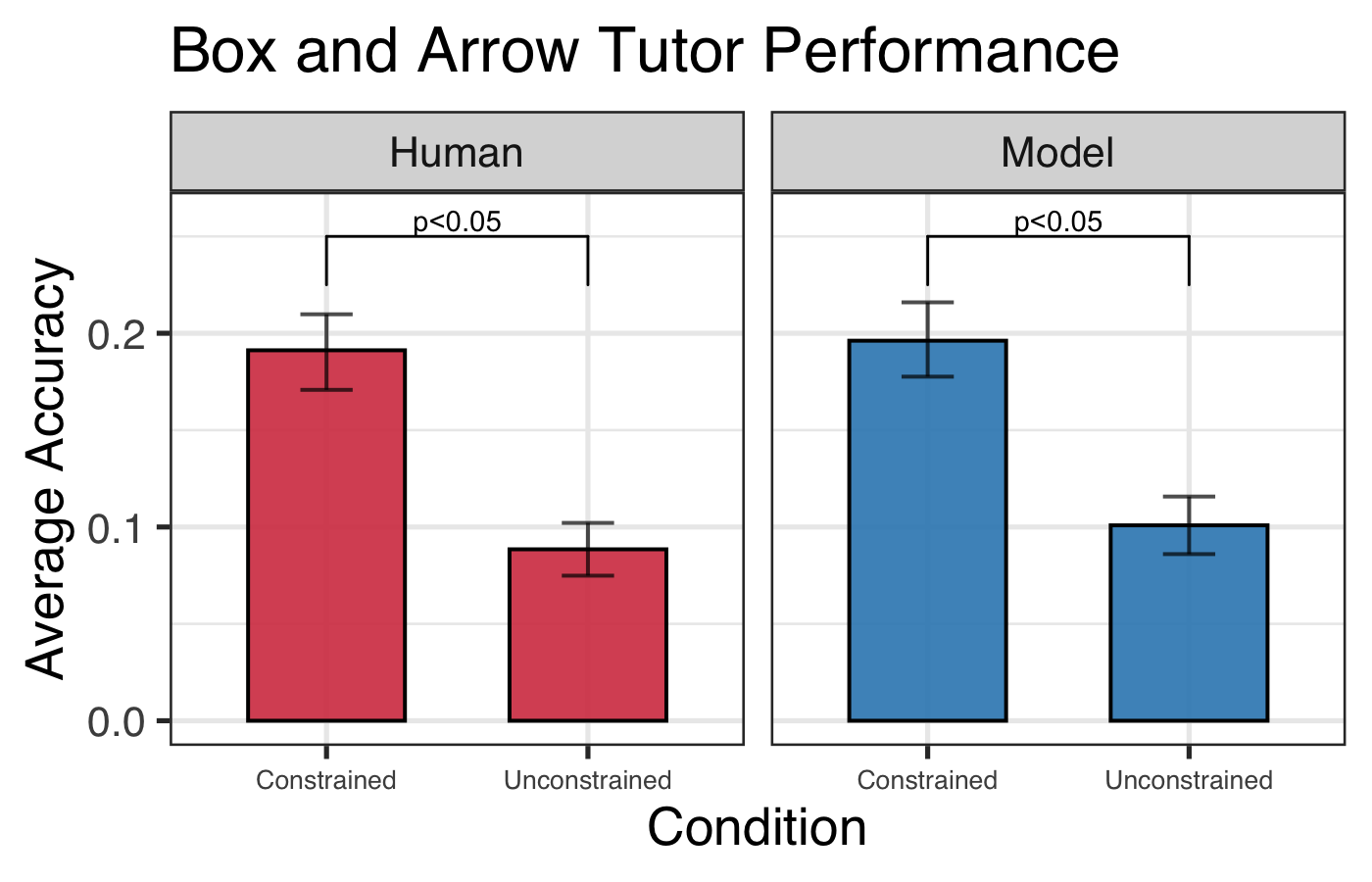}
    \caption{Overall box and arrow accuracy with 95\% CIs.}
    \label{fig:box-and-arrow-overall}
\end{figure}

\begin{figure}[t] 
    \centering
    \begin{subfigure}[b]{0.47\textwidth} 
        \includegraphics[width=\textwidth]{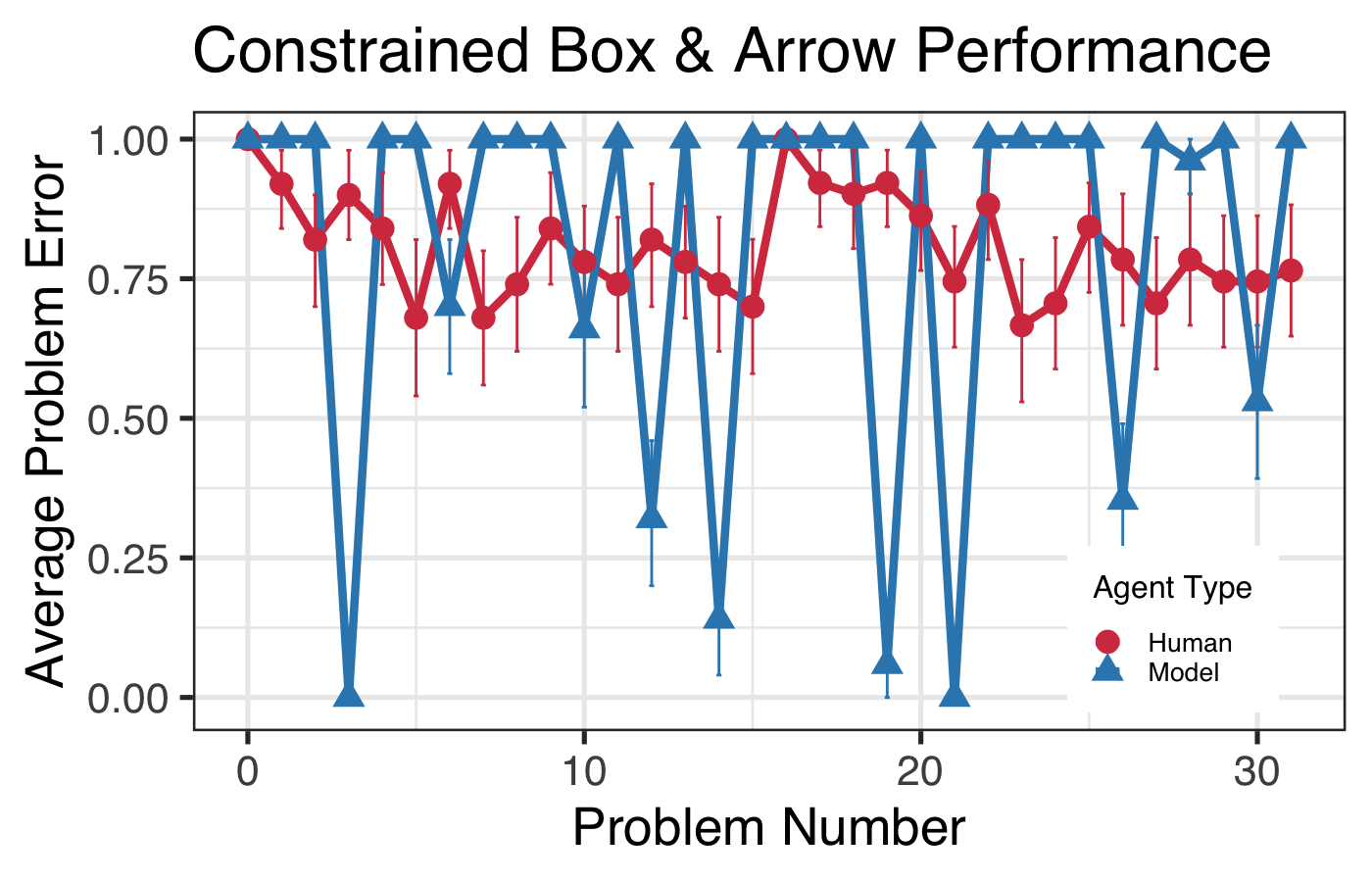}
    \end{subfigure}
    \hfill 
    \begin{subfigure}[b]{0.47\textwidth} 
        \includegraphics[width=\textwidth]{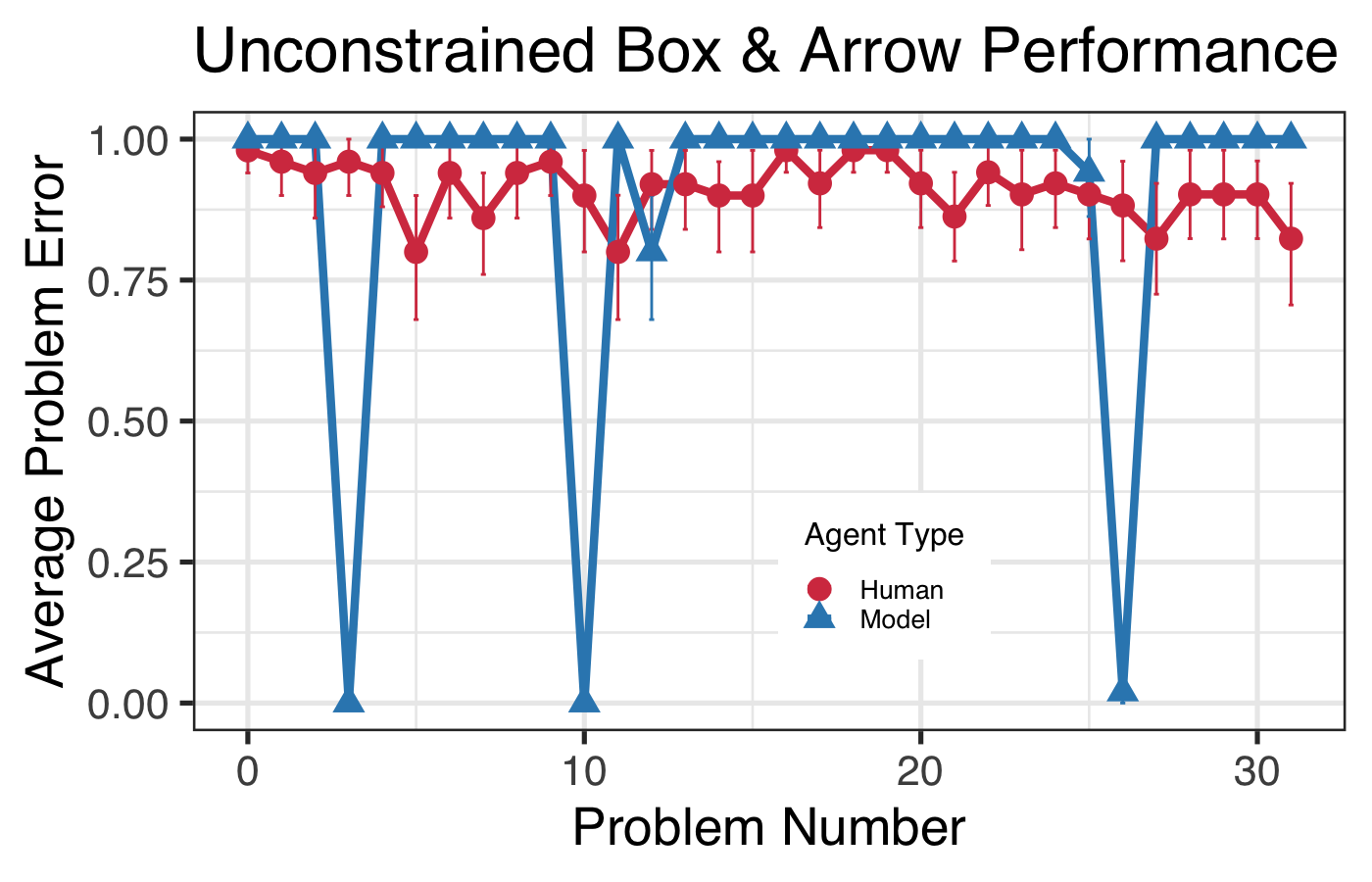}
    \end{subfigure}
    \caption{Box and arrow learning curves with 95\% CIs.} 
    \label{fig:box-and-arrow-learning-curve}
\end{figure}

\subsubsection{Main Effect of Condition}
Figure~\ref{fig:box-and-arrow-overall} shows the accuracy of humans and agents across the two conditions.
Similar to the fractions study, it shows the model and human performance are {\it qualitatively} similar.
I find that \citeA{Lee:2015gb}'s findings apply to just the hard problems: students in the unconstrained condition are less likely to be correct (odds ratio: 0.17, 95\% CI: [0.08, 0.37], $p<0.05$).
I find that agents exhibit the same effect: those trained with unconstrained problems are also less likely to be correct (odds ratio: 0.46, 95\% CI: [0.37, 0.56], $p<0.05$).

This task is also uniquely well suited for comparing the {\it quantitative} performance between agents and humans.
In contrast to the fractions task, where humans could apply prior knowledge, this task was intentionally designed to inhibit prior knowledge use.
Thus, the agents and humans start with the same initial conditions.
Figure~\ref{fig:box-and-arrow-overall} shows that humans have a constrained accuracy of $19.1\%$ ($SD=0.39$) and agents have a similar constrained accuracy of $19.6\%$ ($SD=0.40$). Also, humans have an unconstrained accuracy of $8.8\%$ ($SD=0.28$) while agents have an unconstrained accuracy of $10\%$ ($SD=0.30$).
\vskip 6pt

\subsubsection{Learning Curves}
Figure~\ref{fig:box-and-arrow-learning-curve} shows the learning trajectories of the agents and humans, averaging error across students within each condition at each problem.
In contrast to the fractions results, we see a much shallower rates of learning for both the agents and humans, suggesting that the model is able to capture general trends in learning (rapid learning in fractions, but more limited learning here).
This is notable because the model is not trained on or fit to the human data; these predictions are generated based entirely on the structure of the task and the sequence of the items.
They could have been generated prior to collecting any human data.

\subsection{Discussion}

These findings provide a second demonstration of the model accurately predicting the main effect of a human experiment.
They also showcase again its ability to generate learning curves that reasonably align with human learning trajectories.
Although the study 1 and 2 tasks are both math related, they employ different instructional manipulations---a problem ordering manipulation and an item design manipulation.
This provides some evidence for the model's general ability to guide instructional design for varying kinds of interventions.

The model can also facilitate testing of different learning theories and hypotheses.
My model directly instantiates \citeA{Lee:2015gb}'s  theory of learning as search through a space of rules; its ability to generate data that matches the human data provides additional evidence in support of this theory.
Surprisingly, however, the results provide evidence against \citeA{Lee:2015gb}'s hypothesis that constrained problems aid learning by making the correct procedures easier to compute.
Computing with whole or fractional numbers is equally difficult for agents, but they still exhibits the main effect, so constrained problems must aid agent learning in another way.

I hypothesize that the benefit of constrained problems is less about ease of computation and more about lower procedural ambiguity. 
To satisfy the property that only correct procedures yield whole-number solutions, constrained problems often have only a single (correct) candidate procedure that is consistent with the correct answer.
In contrast, unconstrained problems often have multiple candidate procedures (some incorrect) that are consistent.\footnote{For example, the unconstrained hard problem in Figure~\ref{fig:box-and-arrows-tutor} has three candidates: $7-3=4$ (correct), $2+2=4$, and $2 \times 2=4$. The comparable constrained problem (see Figure 3 in \citeNP{Lee:2015gb}) has only a single (correct) candidate procedure.}
As a result, both the model and humans are more likely to select the correct procedure on constrained problems, which aids their learning and performance.
More research is needed to test this hypothesis, but if true, it would suggest a distinctively different strategy to designing problems to promote learning---items should be designed so there is only a single (correct) candidate procedure that is consistent with the correct answer.



\section{Conclusion}
To my knowledge, this paper presents the first evidence that a computational model of learning can successfully predict the main effects observed in multiple human A/B experiments.
It also demonstrates that this model can generate reasonable predictions of human learning trajectories and offer theoretical insights into the effectiveness of specific instructional interventions.
These findings suggest that computational models of learning could operate as Model Human Learners, supporting instructional designers in evaluating and identifying the most pedagogically effective instructional designs.
More research is needed to expand the Model Human Learner concept to additional tasks and interventions, but this work represents a preliminary proof of concept that lays the groundwork for a broader research program.

\section{Acknowledgments}
This work grew out of my PhD dissertation \cite{Maclellan:2017thesis}. As a result, it would not have been possible without the members of my thesis committee (Ken Koedinger, Pat Langley, Vincent Aleven, and John Anderson) who provided me with invaluable feedback, guidance, and support. This work has been generously funded in part from several sources, including a Graduate Training Grant awarded to Carnegie Mellon University by the Department of Education (\#R305B090023 and \#R305A090519), by the Pittsburgh Science of Learning Center, which is funded by the NSF (\#SBE-0836012), two National Science Foundation Awards (\#DRL-0910176 and \#DRL-1252440), a DARPA award (\#HR00111990055), and the NSF National AI Institutes Program (\#2247790 and \#2112532). The views, opinions and/or findings expressed are those of the author and should not be interpreted as representing the official views or policies of the Department of Defense or the U.S. Government.  I would also like to thank Rony Patel, Hee Seung Lee, and the PSLC DataShop team for providing access to the data used in this paper. Finally, this work would also not have been possible without my collaborator Erik Harpstead, who collaborated with me to first articulate the idea of a computational model of human learning in our original paper on the topic \cite{MacLellan:2016tqa}.

\bibliographystyle{apacite}

\setlength{\bibleftmargin}{.125in}
\setlength{\bibindent}{-\bibleftmargin}

\bibliography{references}

\end{document}